\documentclass[final,3p,times]{elsarticle}

\usepackage{graphicx}% Include figure files
\usepackage{lineno}
\usepackage{amsmath}
\usepackage{ulem}
\usepackage{color}

\usepackage{bm}% bold math

\usepackage{soul}

\begin{document}

\title{Wealth distribution on a dynamic complex network}

%\preprint{APS/123-QED}
\author{G. L. Kohlrausch}
\author{Sebastian Gonçalves}
\address{Instituto de Física, Universidade Federal do Rio Grande do Sul,  91501-970 Porto Alegre, RS, Brazil}%Lines break automatically or can be forced with \\

\begin{abstract}

We present an agent-based model of microscopic wealth exchange in a dynamic network to study the topological features associated with economic inequality. The model evolves through two alternating processes, the conservative exchange of wealth between connected agents and the rewiring of connections, which depends on the wealth of the agents.
The two dynamics are interrelated; from the dynamics of wealth a complex network emerges and the network in turn dictates who interacts with whom.
We study the time evolution and the economic and topological asymptotic characteristics of the model for different values of a social protection factor $f$, which favors the poorest agent in each wealth transaction. In the case of $f=0$, our results show condensation of wealth and connections in a few agents, in accordance with the mean field models with respect to wealth. Low values of $f$ favor agents from the middle and upper classes, leading to the formation of hubs in the network. As $f$ increases, the network restriction on exchanges gives rise to an egalitarian society different from the results outside the midfield network.
\end{abstract} 
\maketitle

\section{Introduction}
Economic inequality is a growing problem of modern societies. Indeed, income and wealth disparities have increased substantially since the beginning of the  XXI century~\cite{Alvaredo2018, CapitalXXI,IncomeChancel,saez2016wealth}.
Despite the geographical, cultural, and historical differences, income distribution in different countries follows the same pattern~\cite{Chakrabarti2013} with mostly a two-class society division~\cite{Yakovenko2001}. The wealthiest class, usually inferior to $5\%$ \cite{Clementi2005}, presents a power-law distribution, and the poorer class, follows a Gamma distribution.
This universal behavior indicates that there must be a set of fundamental mechanisms that lead to economic inequalities regardless of the countries. Understanding this basic mechanism that creates such stratification is essential to devise efficient ways to avoid or reduce it. 
Nevertheless, economic phenomena are complex, with many processes at play simultaneously that are difficult to isolate and assess independently in real societies. 

In this sense, one simple and convenient way of doing that search is to simulate a virtual society where we can set the rules of interaction between economic agents and observe its influences on the macroscopic quantities. Following this idea, a great variety of agent-based models have been recently studied \cite{Benhur, Nener2021,Iglesias2021,Liu2021,Yakovenko2009,Laguna2021, Chakraborti2000, Chatterjee2004, Li2019}. The main strength of this approach is how easily different factors can be incorporated into the system, such as taxation rules \cite{Iglesias2021}, economic growth \cite{Liu2021}, and rationality \cite{Laguna2021}. 

On the other hand, a growing and promising field of study in economics is complex network theory \cite{Bardoscia2021, liu2022preferential, liu2021scale, de2022impact}.
In this approach,  financial institutions or economic agents are the network's nodes, and relations among them are the edges.
However, to build a complete financial network, there are operational problems such as the size of the system and the lack of information on the interconnections between financial institutions, information which is no generally available. In this way, statistical physics approaches have been proposed to reconstruct the real network using the available information \cite{Cimini2019, Squartini2018}. 
This task is even more complicated when we know that the actual connections change over time, thus, the entire network changes.

So far, there are few contributions which includes complex networks into agent-based models  \cite{Ma2013, Braunstein2013}, and even in those, the network topology is fixed in time. In this way, our goal is to provide a framework where a dynamic complex network is incorporated into an agent-based model, enabling a topological analysis of economic inequality phenomena. To study the disparities arising from economic transactions, we consider a conservative market, thus excluding the influences of processes such as the production of wealth and capital appreciation. 

To create a dynamic network related to the wealth exchange process, we propose a model that considers that an agent's degree depends on its wealth. We justify this idea by considering that wealthier financial institutions can create more diversified investment portfolios. In the same way, wealthier companies or industries can reach a more significant number of investors or consumers. Thus, we propose a model which alternates between two dependent processes: the exchange of wealth between connected agents and the rewiring of the network connections.

We describe the model in Sec.~\ref{secmodel}, the results in Sec. \ref{secresults} and in Sec. \ref{secconclu} we present our conclusions. 

\section{Model}\label{secmodel}
The dynamics of the model starts by attributing to each of the agents a wealth $\omega$, which is a random value in the $[0,1)$ interval, and a saving fraction $\alpha$.
Then, we select $z$ agents and fully connect them as the initial condition of the network. After that, we select a random unconnected agent $j$ to connect it to a network agent $i$ with a probability
\begin{equation}\label{prob_con}
 P_{\textrm{connection}}^{i,j} = \frac{\omega_i(t) +\omega_j(t)}{\sum_l \omega_l(t)},
\end{equation}
where the sum in $l$ is only on agents with at least one connection and $\omega_i(t)$ is the wealth of agent $i$ at time $t$.
For each newly added agent $j$, we try the connection with all the already connected  agents, restraining $j$ to at most $z$ connections. This process continues until the $N$ agents are added to the network.

%{\R We set $z=3$,  but this limit does not interfere with the dynamics because it is only used to initiate the network.}\\
%{\B Essa frase é confusa. Se refere a que o z inical é arbitrario, com em B-A? Se for isso não mencionaria aqui, e nao com esse termos (limit?) }\\
After the network is set up, we begin two alternating processes, the exchange of wealth between the connected agents and the rewiring of the connections. 

In the wealth exchange process, each agent trades wealth with all its connections following the yard-sale rule  \cite{Benhur, Hayes}, which states that the amount of wealth traded between agents $i$ and $j$ is 
\begin{equation}
    d\omega = \textrm{min}[\alpha_i\omega_i(t);\alpha_j\omega_j(t)]. 
\end{equation}
%% seguir
To determine in which way the wealth will flow, we define a probability of the poorer agent winning the transaction, which is given by \cite{Scafetta}
\begin{equation}\label{prob_exc}
     P_{\textrm{exchange}}^{i,j} = \frac{1}{2} + f \times \frac{|\omega_i(t)-\omega_j(t)|}{\omega_i(t)+\omega_j(t)},
\end{equation}
where $f$ is the social protection factor, which varies between $0$ and $1/2$. The wealth of the agents $i$ and $j$ after the trade will be 
\begin{align*}
 \omega_i(t+1) = \omega_i(t) + d\omega &&  \omega_j(t+1) = \omega_j(t) - d\omega,
 \end{align*}
where the agent $i$ is the winner of the transaction. In this way the total wealth is always conserved. After all the connected agents have traded wealth with all their connections we start the rewiring process of the network.

The rewiring process starts by randomly selecting a pair $i',j'$ of agents, if this pair is disconnected the probability of creating a new connection follows Eq.\ref{prob_con}, if the pair is already connected then there is a probability $Q = 1 - P_{\textrm{connection}}^{i',j'}$ of breaking the connection. In order to all the agents being able to participate in this process, we select $N/2$ pairs of agents to rewire their connections. After this process is complete, we go back to the wealth exchange process respecting the new network that has been formed. In this way, the evolution of the system depends on the wealth exchange and rewiring processes, which are not independent. 

We use a MCS (acronym for {\it Monte Carlo Step}) as the unit of time, which is defined as the time necessary to make all the wealth exchanges and rewire the $N/2$ connections. We perform simulations for systems with size $N=10^3$ which evolve for $t=4\cdot10^4 MCS$. The results are averaged over $10^3$ independent samples.  
 
\section{Results}\label{secresults}

\subsection{Economic quantities}

We start the analysis of the results by looking at the economic indicators and comparing then with the mean-field model. In this sense, one of the most common measures of economic inequality used by economists, countries and statistical organizations is the Gini index, which measures the statistical dispersion in the wealth distribution. The Gini index can be defined in a operational way as \cite{Sen}
\begin{equation}\label{gini_eq}
    G(t) =\frac{1}{N} \frac{\sum_i^N \sum_j^{i-1}|\omega_i(t)-\omega_j(t)|}{\sum_i \omega_i(t)}.
\end{equation}   
This index varies between $0$, which corresponds to perfect equality among the agents, and $1$ that represents the total concentration of wealth in one agent. 

\begin{figure}
    \centering
    \includegraphics[width=0.5\columnwidth]{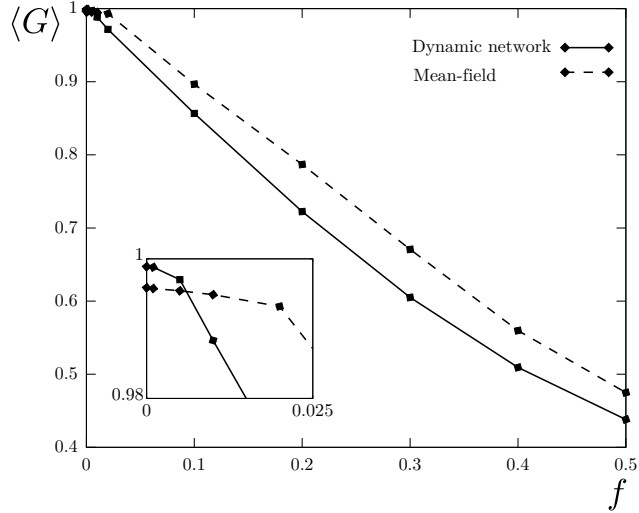}
    \caption{Gini index as a function of the social protection factor for the dynamic network (solid line) and mean-field (dashed line) model. Inset: zoom in the region for  $f=[0:0.025]$.}\label{gini_fig}
\end{figure}

In Fig~\ref{gini_fig} we present the results for the Gini index comparing the present results with the mean-field model results. As we can see, the inclusion of a dynamic network in the model leads to lower values of the Gini for almost all values of $f$. However, when $f\rightarrow 0$ both models lead to the condensation of wealth in one or a few agents. Nevertheless, with values of $f$ close to zero (Fig~\ref{gini_fig}, inset), the results of the Gini index in the two cases are not the same. While in the network model, the Gini index continuously decreases as $f$ increases, in the mean field model, the Gini index remains at Gini $\approx 1$, in a narrow region of values of $f$ close to zero. This could indicate that the two models behave differently for $f \approx 0$ to zero.
Therefore, to try to understand the origin of the differences between the two models, we look to the accumulated distribution of wealth in three distinct regimes: no social protection ($f=0.0$),  small protection ($f=0.01$) and large social protection ($f > 0.1$).

\begin{figure}
\centering
    \includegraphics[width=1\columnwidth]{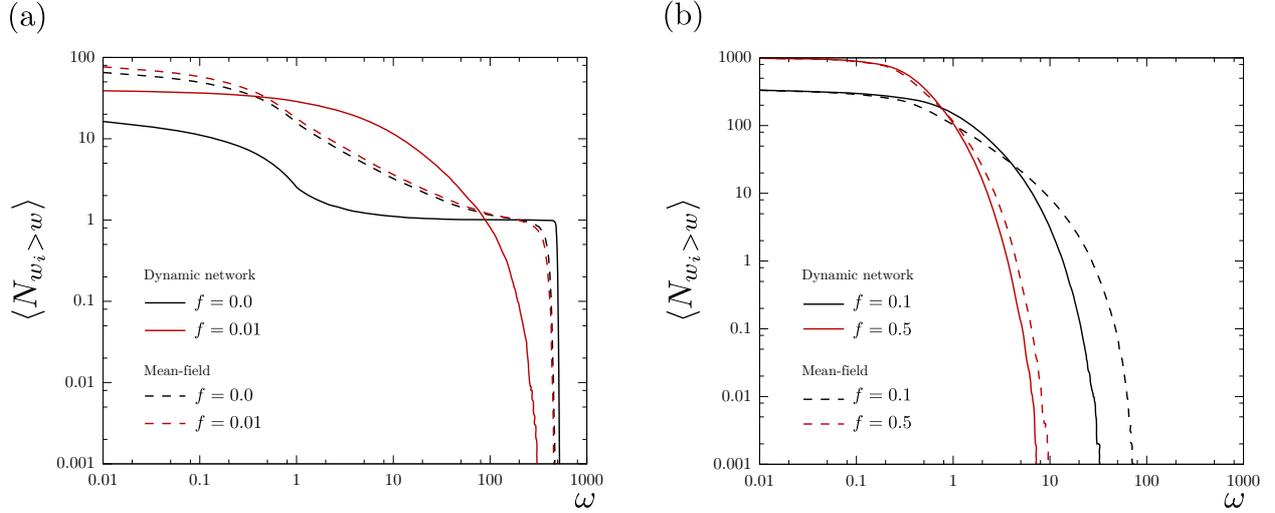}
    \caption{Accumulated distribution of wealth for (a) low and (b) high values of social protection in the dynamic network (solid lines) and mean-field (dashed lines) model.}\label{accum_wealth}
\end{figure}

In Fig~\ref{accum_wealth} we present the accumulated distribution of wealth for different values of $f$, which represents the number of agents whose wealth is greater than a certain value $\omega$ ($\langle N_{\omega_i > \omega}\rangle$). As the results are presented in log-log scale, this distribution doesn't show agents who posses wealth less than $0.01$. Looking at Fig~\ref{accum_wealth} (a) for $f=0.0$ we notice that even the Gini index is very approximate for the two models, the distribution of wealth is vastly different. In the mean-field model, $\langle N_{\omega_i > \omega}\rangle$ continuously decays as $\omega$ increases, until we reach the maximum wealth ($\omega_{max}\approx500$) where there is an abrupt decay in the distribution. Meanwhile, in the network model there is a {\it plateau} in the distribution which goes from $\omega\approx1$ to $\omega\approx500$. The presence of this {\it plateau} indicates a very distinct separation of classes which doesn't seems to be so  pronounced in the mean-field model.  

The $f=0.01$ results shows that the mean-field model is much more resistant to the social protection factor than the dynamic network model.  In the mean-field case we perceive just a slight change in the distribution,  while in our model $\langle N_{\omega_i > \omega}\rangle$ is  utterly different. For instance, the {\it plateau} present in the $f=0.0$ case vanishes as the  number of agents in the middle class ($\omega=[1,100]$) substantially increases. However, the Gini index in this case remains very close to one, indicating that low values of social protection are able to recover the middle class, but are not able to reach the poorest agents. In addition, for low values of $f$, the quantity of agents with no wealth ($\omega<0.01$) is way greater in the dynamic network than in the mean-field model, which can be seen in the beginning of the distributions. This suggests that the network constraint can exclude the poorer agents from the wealth exchanges process. 

In Fig~\ref{accum_wealth} (b) we see that for great values of $f$ the two models slowly converge to similar distributions. The wealth of the richest agent, which is seen by the end of the distribution, significantly diminishes as $f$ increases ($\omega_{max}\approx30$ and $\omega_{max}\approx75$ for $f=0.1$,  $\omega_{max}\approx7$ and $\omega_{max}\approx10$ for $f=0.5$, for the dynamic network and the mean-field model, respectively). Another important point to notice is the vast number of agents who posses no wealth in the $f=0.1$ case, in contrast with the $f=0.5$ case, where all agents have some wealth. In this way, both models agree that only strong social protection factors are able to protect agents belonging to the bottom classes.

\begin{figure}[!h]
\centering
    \includegraphics[width=1\columnwidth]{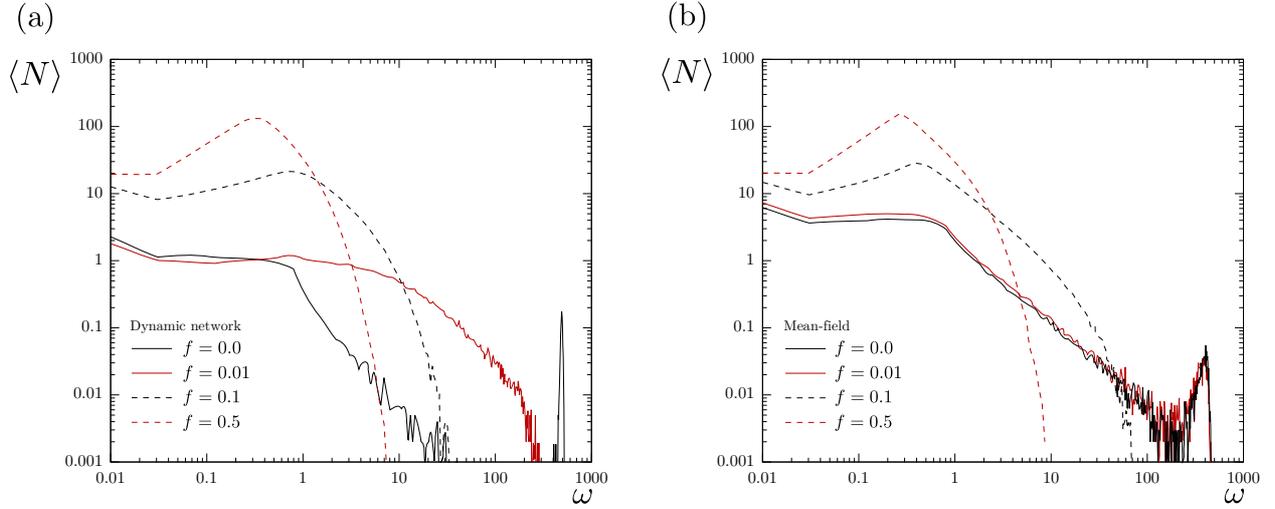}
    \caption{Non-accumulated distribution of wealth for different values of social protection in the (a) dynamic network and (b) mean-field model.} \label{agent_histo}
\end{figure}

In Fig~\ref{agent_histo} we present the mean number of agents per wealth, i.e. the non accumulated distribution of wealth, for (a) the dynamic network and (b) the mean-field model. With this result, it becomes clear that there is a distinct separation of classes for $f=0.0$ in our model, represented by a gap in the distribution of wealth. Once more, a small social protection  can break this scenario, transferring the wealth from the high class to the middle one, as the distribution for the low class is virtually unchanged. The mean-field model also present a separation of classes for $f=0.0$, which can be seen by the three different behaviors in the distribution, but is not as accentuated as in the dynamic network model. Nevertheless, small values of $f$ maintains the distribution of wealth almost unchanged in the mean-field case.

For high values of $f$, in both models, the most noticeable feature in the distribution of wealth is a round maxima that appears at lower values of $\omega$ and greater number of agents as $f$ increases. As the total wealth is conserved and the number of agents with no wealth goes to zero when $f\rightarrow0.5$, there's must be a flow of wealth from the high and middle classes to the bottom one. This flow of wealth recovers the agents who once had no wealth, thus  dislocating the round maxima towards lower values of $\omega$. Also, this round maxima appears in greater values of $\omega$ in the dynamic network than in the mean-field model. Actually, the values in the dynamic network model are $\omega\approx0.8$ and $\langle N \rangle \approx 21$ for $f=0.1$, $\omega\approx0.37$ and $\langle N \rangle \approx 132$ for $f=0.5$, in the mean-field model,  $\omega\approx0.4$ and $\langle N \rangle \approx 28$ for $f=0.1$, $\omega\approx0.27$ and $\langle N \rangle \approx 156$ for $f=0.5$. Here, is important to remark that even a very strong social protection doesn't completely eliminates the economic inequalities in the system. However, it increases the participation of the bottom classes in the wealth exchange process, especially when the network is introduced. In other words, in the presence of a robust social protection, the network constraint on wealth exchanges helps to prevent fast-track gains of wealth that can exclude the poorer agents. To better understand how this constraint acts in the agents wealth we will now present the network measures for our model.

\subsection{Topological quantities}

\begin{figure}[!h]
\centering
    \includegraphics[width=0.5\columnwidth]{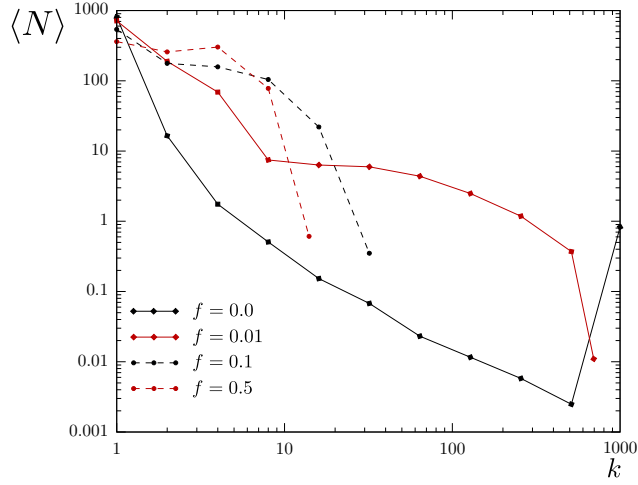}
    \caption{Degree distribution for different values of social protection.} \label{agent_conex}
\end{figure}

We start this analysis by looking at the network degree distribution for different values of $f$ (Fig~\ref{agent_conex}) \footnote{It is important pointing out that in the following results concerning the agents degree ($k$), to avoid noisy distributions in high values of $k$, we grouped the agents in intervals that increase as $2^x$, where $x=[0;10]$. In this way, results for $k=1$ represents agents who posses zero or one connection, the subsequent interval presents agents with two, the next one with three or four, and so on until we reach the maximum value of $k$.}. In the absence of social protection we see a concentration of connections in one agent,  $\langle N \rangle \approx 1$ for $k=999$.  An important point to notice for $f=0.0$ is the values of $\langle N \rangle$ presented for $16<k<999$, in this region $\langle N \rangle < 0.1$. As the results are mean values between $1000$ independent samples, agents in this degree region happened on less than $10\%$ of the samples, meaning that we can consider, without loss of generality, that the network condensates in only one agent. When a weak social protection is introduced into the system, the concentration of connections disappears and the number of highly connected nodes ($k>10$) increases. In this regard, $f=0.01$ dissolves an one agent centered network into a more distributed one, where highly connected agents represent hubs in the network. As $f\rightarrow0.5$ the disparities in the degree distribution diminishes and the maximum value of $k$ is greatly reduced, as a matter of fact, $k_{max} = 999$, $695$, $32$ and $14$ for $f=0.0$, $0.01$, $0.1$ and $0.5$, respectively. 

\begin{figure}[!h]
\centering
    \includegraphics[width=0.5\columnwidth]{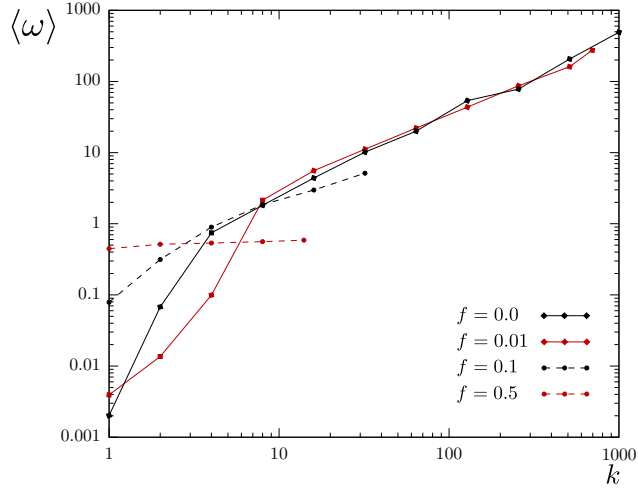}
    \caption{Mean wealth of agents with $k$ connections for different values of social protection.} \label{wealth_conex}
\end{figure}

In order to investigate how the degree and wealth distributions are related we present in Fig~\ref{wealth_conex} the wealth-degree distribution, where we show the mean wealth of agents with $k$ connections. The condensation of wealth and connections when $f=0.0$ happens on the same agent, when $k=999$, $\langle \omega \rangle \approx 500$ and $\langle N \rangle \approx 1$. There is one agent who posses almost all the wealth and is linked to every other one in the network. 
The wealth-degree distributions for $f=0.0$ and $f=0.01$, besides looking very similar for the highly connected agents, strongly differ in the endpoints, for $f=0.01$,  $\langle \omega \rangle \approx 275$ and $k=695$, so the wealth of the most connected agent when $f=0.01$ is nearly half of when $f=0.0$. Also, this two distributions are significantly unlike in smaller values of $k$. To explain this change we look again to Fig~\ref{agent_conex} and perceive that the number of agents in this region ($k=[2,10]$) vastly increases from $f=0.0$ to $f=0.01$. Thereby, when there is no social protection only a few agents who manage to gain a little more wealth, notice the low values of $\langle \omega \rangle$, can become more connected. Otherwise, in the $f=0.01$ case the poorer agents are able to creates more links in the network, thus lowering the mean value of $\omega$ for low values of $k$.

Going through higher values of $f$ we see that the mean wealth still has a strong dependence on the agents degree when $f=0.1$, which remains present, but not so remarked, when $f=0.5$. This way, the agents degree is a determining feature of its wealth, which  determines its degree in the same sense of the opposite. In this manner, wealthier agents are able to allocate their resources in more transactions, again, increasing their profit. This mechanism, which gradually excludes the poorer agents, is greatly attenuated by the increase of the social protection. Another important point is the increase of wealth on the less connected agents ($k=1$) as $f$ increases, it goes from $2\cdot 10^{-3}$ for $f=0.0$ to $4\cdot10^{-3}$ for $f=0.01$, then $7.9\cdot 10^{-2}$ for $f=0.1$ and $0.44$ for $f=0.5$.  

Aiming to gather information on how the agents are connected with each other, our next step will be investigating the network assortativity ($r$), which measures the Pearson correlation of the degree of connected nodes \cite{Newman2002}. This coefficient can be expressed in a operational way as

\begin{equation}
  r = \frac{M^{-1}\sum_i^M j_i k_i - [M^{-1}\sum_i^M \frac{1}{2}(j_i +k_i)]^2}{M^{-1}\sum_i^M \frac{1}{2}(j_i^2 +k_i^2) -[M^{-1}\sum_i^M \frac{1}{2}(j_i +k_i)]^2 },
    \end{equation}

where the sum in $i$ is made on the edges of the network, $M$ is the total number of connections, $j_i$ and $k_i$ are the degrees of the nodes (agents) at the end of the $i$ edge. This coefficient varies in the interval $-1\leq r \leq 1$. For $r<0$ the network is called disassortative and represents a network in which the highly connected nodes are linked to the ones with low degrees. For $r>0$ the network is assortative and the nodes are connected with others with similar degree. Networks with $r=-1$ ($r=1$) are called perfectly disassortative (assortative), when $r=0$ the network is non-assortative.  

\begin{figure}
\centering
    \includegraphics[width=1\columnwidth]{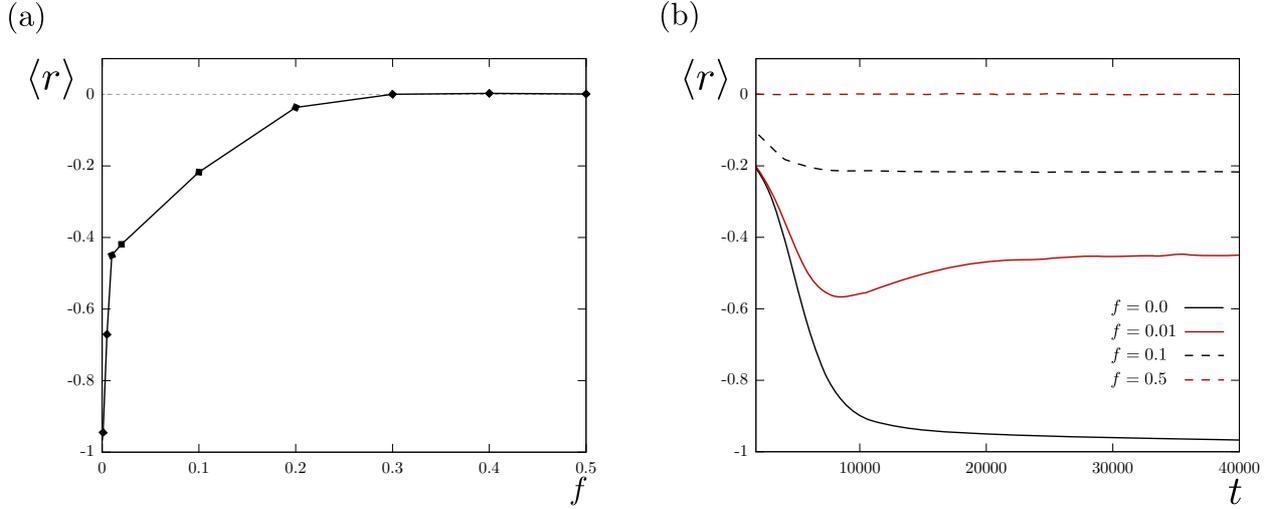}
    \caption{Assortativity of the network as a function of (a) social protection $f$ and (b) time $t$ for different values of $f$.} \label{fig_assor}
\end{figure}

In Fig~\ref{fig_assor} (a) we present the assortativity of the network for different values of social protection. For $f=0.0$ the network is almost perfectly disassortative, $\langle r \rangle = -0.967$, which is in agreement with our previous results indicating a single agent concentrating almost all the connections. As the social protection increases, $|\langle r \rangle|$ rapidly decreases, we found $\langle r \rangle = -0.67$ for $f=0.005$ and $\langle r \rangle = -0.45$ for $f=0.01$. Thus, the assortativity confirms our previous results pointing out that even a very small social protection can break the condensate scenario. As $f\rightarrow0.5$ the network becomes non-assortative, and there is no correlation between the degree of connected agents. 

To evaluate how the network evolves and if our simulations are reaching a steady state, we present in Fig~\ref{fig_assor} (b) the change in the assortativity over simulation time. As shown, $\langle r \rangle$ does not present significant changes after $3\cdot10^4$ MCS for all the values of $f$ presented. Therefore, we can consider that our results using $4\cdot10^4$ MCS are all over a stable network. Furthermore, the network is able to reach a steady state faster for strong values of $f$ than for small ones. It it important to point out that a steady state does not mean that there are no microscopic changes in the network, i.e. rewiring of connections and exchanges of wealth, but that the macroscopic quantities are stable. For $f=0.0$ and $f=0.01$  the network evolves very similarly for a brief period of time, after that, there is a turning point around $1\cdot10^4$ MCS, where $|\langle r \rangle|$ decreases for $f=0.01$. This behaviour indicates that the network was proceeding to the condensate state, yet the small social protection was able to prevent it, leading to the formation of hubs in the network, explaining the high values of the Gini index and the very unequal wealth distributions in this case. 

\begin{figure}
\centering
    \includegraphics[width=0.5\columnwidth]{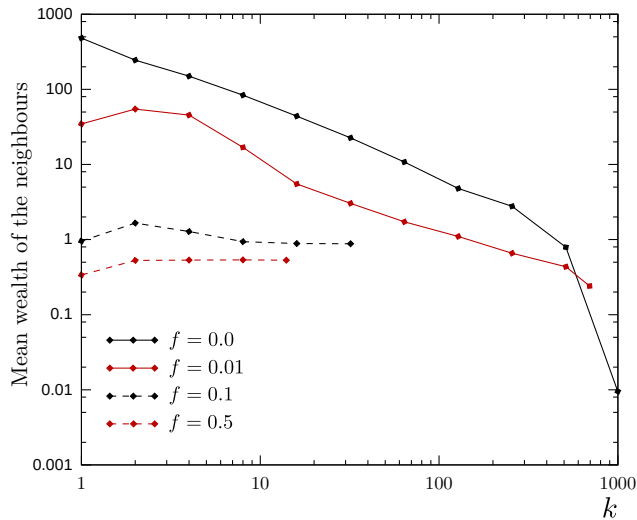}
    \caption{Mean wealth of the neighbours of an agent with $k$ connections for different values of social protection.} \label{neighbours}
\end{figure}

To round off the presentation of our results, we now look at the mean wealth of the neighbours of an agent with degree $k$ (Fig~\ref{neighbours}). For $f=0.0$, the agents who posses one connection are linked with those with wealth near $500$, and agents with $999$ connections have neighbours with mean wealth of $0.01$. This result is in agreement with those presented in Figures \ref{agent_conex}, \ref{wealth_conex} and \ref{fig_assor}.  Moreover, as the wealthier agent is connected with all the other ones, his wealth is always accounted in the mean value,  what explains the linear decay of the mean wealth of the neighbours with the degree. When $f=0.01$, the neighbours of the highly connected nodes are the poorer in the system, meaning that the agents with high degree are able to extract wealth from their neighbours. In particular, the wealthier agents are those connected with agents with $k=2$ and $4$,  looking again to Fig~\ref{wealth_conex},  we can see that these are the poorer agents in the system, so the highly connected and richer agents are those connected to a high number of low connected and poor agents. 

In the case of strong social protections we see that the minimum happens for $k=1$, the maximum for $k=2$, then, decreases when $f=0.1$, and remains constant when $f=0.5$. To explains this behavior is important to observe the assortativity in those cases. For $f=0.1$ the network is slightly disassortative ($\langle r \rangle = -0.217$), in this sense, agents with low degree are more connected with agents with higher values of $k$. However, connections of agents with similar degrees are more common in this case when compared with the small social protection scenario. In this way, if two agents with $k=1$ are linked, they cannot extract much wealth from each other, as they usually are the poorer agents (see Fig~\ref{wealth_conex}). Thus, besides the connections with agents with higher degree substantially increase the mean wealth of the neighbours for $k=1$, the attachment to others with only one connection can explain the minimum happening for this value in both $f=0.1$ and $0.5$. Furthermore, the disassortative behavior for $f=0.1$ helps to clarify why the wealth of the neighbours decreases for $k>2$, as in this case the high degree agents interact more with the low degree ones than with the also highly connected. After all, the non-assortative behavior of the network when $f=0.5$ explains why the mean wealth of the agents remains constant after $k=2$. In this scenario, the agents are connected with other ones without any correlation with their degree, and nevertheless, as shown in Fig~\ref{wealth_conex}, the wealth of an agent does not have a significant dependence with his degree for $f=0.5$.     

\section{Conclusions}\label{secconclu}

Aiming to investigate the microscopic mechanisms that helps to aggravate economic inequalities in modern societies, we introduced an dynamic complex network agent-based model which considers two dependent intermittent processes, the rewiring of connections between agents who acts as nodes of a complex network, and the wealth exchange between connected agents. To define the rewiring process, we consider that one of the fundamental mechanisms of social inequality is that the richer agents are able to allocate their resources in a greater number of different transactions. For an example, one can imagine that wealthier individuals, or financial institutions, are able to divide their wealth in a greater number of assets, creating  more diverse portfolios than the poorer individuals. Following this idea, we introduced a probability of connection that depends on the agents wealth, so a complex network is created favoring the links to wealthier agents. Furthermore, as our goal was to study the economic inequalities which rise from the economic transactions, we consider that the total wealth of the system is always conserved, thus excluding the processes of wealth production and capital appreciation/depreciation.       

In the wealth exchange process, we consider the fair-rule of trade \cite{Hayes,Benhur}, in which the two agents participating in the transactions bet a fraction of their wealth defined by a savings fraction parameter $\alpha$. To define the winner of the transaction we consider a probability of choosing the poorer agent (Eq~\ref{prob_exc}) that depends on a parameter $f$. Which is a social protection of the poorer agents, such as financial regulations or governmental social programs. We investigated the model for different values of $f$ and, to evaluate the influence of the network, compared the economic indicators with mean-field model results. 

%The economic indicators suggested that the dynamic network model could be evaluated in two different regimes, weak ($f=0.0$ and $0.01$) and strong ($f=0.1$ and $0.5$) social protections, which demonstrated a very distinct phenomenology.

The dynamic network and mean-field models lead to similar values of the Gini index in small values of $f$. However, the accumulated and non-accumulated wealth distributions are strongly different. In the absence of social protection both models point out a condensation of wealth. Nevertheless, in our model there is a gap in the wealth distribution, indicating a strong separation of classes, which is not so remarked in the mean-field model. The divergences in these models are even more clear for $f=0.01$, as the mean-field seems to be very robust to small increases of $f$, the dynamic network model presents a very distinct phenomenology from that of $f=0.0$. Howbeit, the divergences presented in our model for small $f$ are concentrated in the middle ($\omega =[1;100]$) and high ($\omega>100$) class of agents, demonstrating that weak social protections are able to favor the middle class of agents but cannot reach the lower class. The network measurements for $f=0.0$ and $0.01$ show that this small social protection breaks a scenario where a single agent concentrates almost all the connections in the network into one where a small set of highly connected and richer agents act as hubs, maintaining a very unequal society in which the poorer agents are peripheral in the network.        

As the social protection factor increases, the distributions of wealth for both models approach each other, yet the dynamic network model leads to smaller values of the Gini index. This results can be explained by the non-assortative behavior of the network as $f\rightarrow0.5$, so the degree of the agents is not correlated with their neighbours. Also, the agents wealth dependence on its degree is strongly reduced as $f$ increases. Thus, this two factors help to approximate the dynamic network to the mean-field model. Nevertheless, the network constraint on wealth exchanges prevents fast gains of wealth, as the agents are limited to interact with their first neighbours, leading to lower values of $G$.   

In summary, our results showed that the dynamic network process has strong consequences in the agent based model, presenting a much richer phenomenology, especially in lower values of $f$. In our opinion, this simple model is able to reproduce distinct topological features which can be related to real societies, such as social stratification and the marginalization of the poorer agents. Furthermore, we believe that more studies in this model, considering different forms of wealth exchange, taxation, redistribution and production of wealth are of great interest.

\section*{Acknowledgments}
This work was supported by the Brazilian funding agencies Conselho Nacional de Desenvolvimento Cient\'ifico e Tecnol\'ogico (CNPq).

\bibliography{Wealth_distribution_on_a_dynamic_complex_network.bbl}

\end{document}